\documentclass[12pt]{article}
\pdfoutput=1
\usepackage{jhep-mod-2018}
\usepackage{bm}
\usepackage{amssymb}
\usepackage{hyperref}
\usepackage{datetime}
\usepackage{pifont}
\newcommand{\xmark}{\ding{55}}
\usepackage[usenames,dvipsnames,svgnames,table]{xcolor}
\def\d{{\mathrm{d}}}

\def\O{{\mathcal{O}}}

\makeatletter
\begin{document}
\title{The Pauli sum rules imply BSM physics}
\author{Matt~Visser}
\affiliation{School of Mathematics and Statistics,
Victoria University of Wellington; \\
PO Box 600, Wellington 6140, New Zealand.}
\emailAdd{matt.visser@sms.vuw.ac.nz}
\abstract{
Some 67 years ago (1951) Wolfgang Pauli mooted the three sum rules:
\[
 \sum_n   (-1)^{2S_n}  g_n = 0;  \qquad  \sum_n  (-1)^{2S_n}  g_n \; m_n^2 =0; 
 \qquad  \sum_n   (-1)^{2S_n}  g_n \; m_n^4=0.
\]
These three sum rules are intimately related to both the \emph{Lorentz  invariance} and the \emph{finiteness} of the zero-point stress-energy tensor. 
Further afield, these three constraints are also intimately related to the existence of \emph{finite QFTs} ultimately based on Fermi--Bose cancellations. (Supersymmetry is neither \emph{necessary} nor \emph{sufficient} for the existence of these finite QFTs; though \emph{softly but explicitly broken} supersymmetry or \emph{mis-aligned supersymmetry} can be used as a book-keeping device to keep the calculations manageable.) 
In the current article I shall instead take these three Pauli sum rules as given, assume their exact non-perturbative validity, contrast them with the observed standard model particle physics spectrum, and use them to extract as much model-independent information as possible regarding beyond standard model (BSM) physics.

\medskip
\noindent
K{\sc{eywords}}:
Pauli sum rules; Lorentz invariance; zero-point stress--energy tensor; \\
zero-point energy density; zero-point pressure;  finite QFTs.


\medskip
\noindent
D{\sc{ate}}:  14 August 2018; 21 August 2018; \LaTeX-ed \today

P{\sc acs:} 04.20.-q; 04.20.Cv; 04.62.+v; 04.70.-s
}

\maketitle
\markboth{The Pauli sum rules imply BSM physics}{}
\section{Introduction}
\def\d{{\mathrm{d}}}

In his ETH lectures of 1951, (transcribed and translated into English in 1971),  
Wolfgang Pauli mooted the three sum rules~\cite{Pauli}
\begin{equation}
 \sum_n   (-1)^{2S_n}  g_n = 0;  \qquad  \sum_n  (-1)^{2S_n}  g_n \; m_n^2 =0; 
 \qquad  \sum_n   (-1)^{2S_n}  g_n \; m_n^4=0.
\label{E:polynomial}
\end{equation}
Here the sum is over all particle species indexed by $n$, 
counting boson contributions as positive and fermion contributions as negative, hence the factor $(-1)^{2S}$. The degeneracy factor $g$ includes a spin factor $g = 2S+1$ for massive particles, whereas the spin factor is $g=2$ for massless particles.  The degeneracy factor $g$ also includes an additional factor of $2$ when particle and antiparticle are distinct, and an additional factor of 3 due to colour. (So for example, $g=2$  for the photon, $g=4$ for the electron, and $g=12$ for quarks.) Finally one sums over all particle species indexed by $n$. It is the physical relevance of this sum over the entire particle physics spectrum, leading to significant bose--fermi cancellations,  that is Pauli's key physical insight. (In related discussion in reference~\cite{Visser:1995} the bose--fermi sign factor $(-1)^{2S}$ has been absorbed into the degeneracy factor $g$.) 
Note that the bose--fermi cancellations coming from the sum over particle species can be thought of as a physical, unitarity preserving, form of Pauli--Villars regularization --- see discussion in reference~\cite{Visser:2016}.

\clearpage
When viewed with hindsight these three Pauli conditions are necessary and sufficient for ensuring both the \emph{Lorentz invariance} and \emph{finiteness} of the zero-point stress-energy tensor~\cite{Visser:2016}. Specifically, (see for instance reference~\cite{Visser:2016} and many references therein), once the three sum rules above are imposed one deduces that the net zero-point energy density and net zero-point pressure are given by the finite expression:
\begin{equation}
\rho_{zpe} = -  p_{zpe} =    {\hbar\over 64\pi^2}  \sum_n   (-1)^{2S_n}  g_n \; m_n^4\;  \ln (m_n^2/\mu_*^2). 
\label{E:finite}
\end{equation}

Historically, Pauli actually imposed a fourth sum rule 
\begin{equation}
 \sum_n   (-1)^{2S_n}  g_n \; m_n^4 \; \ln (m_n^2/\mu_*^2) = 0,
 \label{E:logarithmic}
\end{equation}
in order to set the zero-point energy density \emph{exactly} to zero --- I shall argue that this is unnecessary and unhelpful for current purposes, and we shall allow this quantity to be nonzero. (Nonzero but in some sense small, since this quantity determines the effective cosmological constant.) This fourth logarithmic-in-mass quantity is actually independent of the arbitrary parameter $\mu_*$ as long as one has already imposed the third of the polynomial-in-mass conditions.
(Various comments along somewhat similar lines can be found in references~\cite{Akhmedov:2002, Ossola:2003, Culetu:2004, Kamenshchik:2006, Alberghi:2008, Mannheim:2011, Koksma:2011, Asorey:2012,  Mannheim:2016, Kamenshchik:2016, Kamenshchik:2018, Ejlli:2017}.)
In this current article I will assume the exact non-perturbative validity of Pauli's sum rules and use them to extract --- in a largely model-independent way --- as much information as possible (both qualitative and quantitative) regarding beyond-standard-model (BSM) physics.

\section{Supertrace formulation of the Pauli sum rules}

If desired one can rewrite the sum over the particle spectrum in Pauli's sum rules as a ``supertrace'',
\begin{equation}
\sum_{n}   (-1)^{2S_n}  \,g_n  \, X_n =   \mathrm{Str}[ X].
\end{equation}
This is merely a book-keeping device, it is not \emph{per se} an appeal to supersymmetry. 
(See particularly reference~\cite{Visser:2002} and the more recent extensive discussion in reference~\cite{Visser:2016}. Any of the options of \emph{softly but explicitly broken} supersymmetry, the known non-supersymmetric UV-finite QFTs~\cite{Howe:1983, Parkes:1983, Parkes:1984, Kazakov:1995, Piguet:1996, Kobayashi:1997},  or \emph{mis-aligned supersymmetry}~\cite{Dienes:1994, Dienes:1995a, Dienes:1995b, Dienes:2001} can be used to motivate introducing this book-keeping device.)

Pauli's three polynomial-in-mass sum rules would then be
\begin{equation}
\mathrm{Str}[1] = 0;  \qquad  \mathrm{Str}[m^2] =0; \qquad \mathrm{Str}[m^4]=0.
\label{E:polynomial2}
\end{equation}
Furthermore the net zero-point energy density and net zero-point pressure are then:
\begin{equation}
\rho_{zpe} = -  p_{zpe} =    {\hbar\over 64\pi^2} \; \mathrm{Str}[m^4\;\ln (m^2/\mu_*^2)]. 
\label{E:finite2}
\end{equation}
This book-keeping device is nevertheless extremely useful in terms of simplifying the calculations and the presentation in the discussion below.

\section{Standard model particle spectrum}

Let us now divide the particle physics spectrum into standard model (SM) and beyond standard model (BSM) sectors. 
Pauli's three sum rules can then be rewritten as:
\begin{equation}
\mathrm{Str}_{BSM}[1] = N_{BSM}= 
 -  \mathrm{Str}_{SM}[1];  
 \label{E:bsm1}
\end{equation}
\begin{equation}
 \mathrm{Str}_{BSM}[m^2] = (M_2)^2 =
  -  \mathrm{Str}_{SM}[m^2];  
 \label{E:bsm2}
\end{equation}
\begin{equation}
\mathrm{Str}_{BSM}[m^4]= (M_4)^4 =
  -  \mathrm{Str}_{SM}[m^4].
 \label{E:bsm3}
\end{equation}

Now the standard model particles are by definition ``known'', and we now have quite good estimates for their masses. 
Relevant data (from the PDG, {\sf pdg.lanl.gov}~\cite{pdg}) is presented in Table 1. In that table $d= (-1)^{2S} g$, while $\hat m = m/m_H$ is the dimensionless mass (in terms of the Higgs mass). Note that I treat neutrinos as Dirac particles with both left-handed and right-handed components, this being the minimalist extension of the original standard model to include neutrino masses.

\begin{table}[!h]
\begin{center}
\caption{\bf Particle masses in the standard model of particle physics.}

\medskip
\makebox[\textwidth][c]{
\begin{tabular}{|c|r|c|c|c|c|c|}
\hline
\hline
particle	&	$d$	&	mass/GeV	&	$\hat m = m/m_H$	&  $d\times \hat m^2$ &	$d \times \hat m^4$	&	$d \times \hat m^4 \ln (\hat m^2)$	\\
\hline
\hline
Higgs	&	+1	&	$125.02$	&	1	&	1	&	1	&	0	\\
$Z^0$	&	+3	&	$91.1876$	&	$0.729384099$	&	$1.59600349$	&	$0.849075713$	&	$-0.535859836$\\
$W^\pm$	&	+6	&	$80.385$	&	$0.642977124$	&	$2.480517489$	&	$1.025494502$	&	$-0.905811363$\\
\hline
top	&	$-12$	&	$173.21$	&	$1.385458327$	&	$-23.0339373$	&	$-44.21352229$	&	$-28.82995836$\\
bottom	&	$-12$	&	$4.66$	&	$0.037274036$	&	$-0.016672245$	&	$-2.31636^{-05}$	&	$0.000152392$	\\
charm	&	$-12$	&	$1.27$	&	$0.010158375$	&	$-0.001238311$	&	$-1.27784^{-07}$	&	$1.17292^{-06}$\\
strange	&	$-12$	&	$0.096$	&	$0.000767877$	&	$-7.07562^{-06}$	&	$-4.17204^{-12}$	&	$5.98427^{-11}$\\
up	&	        $-12$	&	0.0022	&	$1.75972^{-05}$	&	$-3.71593^{-09}$	&	$-1.15068^{-18}$	&	$2.51947^{-17}$\\
down	&	$-12$	&	0.0047	&	0.000037594	&	$-1.69597^{-08}$	&	$-2.39693^{-17}$	&	$4.8843^{-16}$	\\
\hline
gluons	&	+16	&	0	&	0	&	0	&	0	&	0	\\
\hline
tau	&	$-4$	&	1.77686	&	0.014212606	&	$-0.000807993$	&	$-1.63213^{-07}$	&	$1.38849^{-06}$\\
muon	&	$-4$	&	0.105658375	&	0.000845132	&	$-2.85699^{-06}$	&	$-2.0406^{-12}$	&	$2.88786^{-11}$\\
electron	&	$-4$	&	0.000510999	&	$4.08734^{-06}$	&	$-6.68253^{-11}$	&	$-1.11641^{-21}$	&$2.77039^{-20}$\\
\hline
neutrinos	&	$-$12	&	0.000000002	&	$1.59974^{-11}$	&	$-3.07102^{-21}$	&	$-7.85929^{-43}$	&	$3.90742^{-41}$\\
\hline
photon	&	+2	&	0	&	0	&	0	&	0	&	0\\
\hline
\hline
Str$_{SM}$[X]	&	$-68$	&	\xmark	&	\xmark	&	$-17.97614482$	&	$-41.33897553$	&	$-30.27147461$\\
\hline
\hline
\end{tabular}
}

\smallskip
{\bf Explanation:} Calculations of Str$_{SM}$[1], Str$_{SM}$[$\hat m^2$], Str$_{SM}$[$\hat m^4$], and Str$_{SM}$[$\hat m^4 \ln \hat m^2$], working in the SM sector after spontaneous electro-weak symmetry breaking. 
\end{center}
\label{T:1}

\end{table}%

From Table 1 is clear that within the SM sector, the three quantities Str$_{SM}$[$\hat m^2$], Str$_{SM}$[$\hat m^4$], and Str$_{SM}$[$\hat m^4 \ln \hat m^2$] are utterly dominated by the top quark --- with the top quark accounting for some 80\% to 95\% of the SM effect. This happens for two reasons, first the top quark is simply the heaviest SM particle, and secondly the degeneracy factor for quarks ($g=12$) is so high. Even if one looks slightly beyond the top quark itself, between them the Higgs, $Z^0$, $W^\pm$, and the top quark are the only particles making any appreciable contribution to these quantities from within the SM sector.
We note
\begin{equation}
N_{BSM} = 68;  
\qquad M_2 =  4.240 \;m_H; 
\qquad M_4 = 2.536 \; m_H. 
\end{equation}

In particular, assuming validity of the first Pauli constraint, one immediately sees that while the SM is fermi dominated the BSM sector is boson dominated --- there are at least 68 bosonic degrees of freedom in the BSM sector. Furthermore the $m^2$ and $m^4$ BSM contributions to the sum rules, which determine $M_2$ and $M_4$, also indicate that the BSM spectrum is boson dominated. 
Indeed we can to some extent quantify this observation by writing:
\begin{equation}
\mathrm{tr}_{BSM}^{bose}[1] = N_{BSM} + \mathrm{tr}_{BSM}^{fermi}[1] \geq N_{BSM};
\end{equation}
\begin{equation}
\mathrm{tr}_{BSM}^{bose}[m^2] = (M_2)^2 + \mathrm{tr}_{BSM}^{fermi}[m^2] \geq  (M_2)^2;
\end{equation}
\begin{equation}
\mathrm{tr}_{BSM}^{bose}[m^4] = (M_4)^2 + \mathrm{tr}_{BSM}^{fermi}[m^4] \geq  (M_4)^2.
\end{equation}

\section{Cosmological constant}

Now define two energy scales, $\mu_{SM}$ and $\mu_{BSM}$, characteristic of the SM and BSM particle spectra, by setting
\begin{equation}
\hbox{Str}_{SM} [ m^4 \ln (m^2/\mu_{SM}^2)] = 0; \qquad \hbox{Str}_{BSM} [ m^4 \ln (m^2/\mu_{BSM}^2)] = 0.
\end{equation}

Then
\begin{equation}
\mu_{SM}^2 = m_H^2 \exp \left(  \hbox{Str}_{SM} [\hat m^4\ln \hat m^2]\over \hbox{Str}_{SM} [ \hat m^4] \right) =
2.080 \; m_H^2 = (1.442)^2 \; m_H^2. 
\end{equation}
Unfortunately we have no similar result for $\mu_{BSM}$; while we know (assuming the second ant third Pauli constraints) both $\hbox{Str}_{BSM} [m^2]$ and $\hbox{Str}_{BSM} [m^4]$, we do not (at this stage) have any information regarding $\hbox{Str}_{BSM} [m^4\ln(m^2/\mu_*^2)]$ --- this aspect of the BSM spectrum is, (at this stage), not all that tightly constrained.

However we do know that the cosmological constant can be estimated by~\cite{Visser:2016} 
\begin{eqnarray}
\rho_{cc}  = \rho_{zpe} = -  p_{zpe} 
&=& - {\hbar\over 64\pi^2} \left\{\hbox{Str}_{SM} [m^4] \right\} \ln \left(\mu_{BSM}^2\over\mu_{SM}^2\right)
\nonumber\\
&=& \quad { 0.06545\; \hbar \; m_H^4} \; \ln \left(\mu_{BSM}^2\over\mu_{SM}^2\right)
\nonumber\\
&=& \quad { (0.50580)^4\; \hbar \; m_H^4} \; \ln \left(\mu_{BSM}^2\over\mu_{SM}^2\right).
\end{eqnarray}
Here $\mu_{BSM}$ is the only place that unknown BSM physics now enters, and only as a single parameter,  into the cosmological constant. 

At least the energy scale for the cosmological constant is not off by the extremely naive factor $10^{123}$; it is now more like $10^{55}$.  Roughly speaking, the $\rho_{cc}\sim\O(M_\mathrm{Planck}^4)$ guesstimate has been replaced by a $\rho_{cc}\sim\O(m_\mathrm{Higgs}^4)$ estimate.  It is important to realise that it is \emph{not} supersymmetry that is responsible for this vast reduction in scale, it is the much more basic symmetry of Lorentz invariance for the zero-point stress-energy~\cite{Visser:2016}. The observational data regarding the cosmological constant now suggests
\begin{equation}
0  \lesssim \ln \left(\mu_{BSM}^2\over\mu_{SM}^2\right) \lesssim 10^{-55}. 
\end{equation}

\clearpage
It is probably best to interpret this as an extremely tight observational (rather than theoretical) constraint on the BSM spectrum. Equivalently
\begin{equation}
0\lesssim  \hbox{Str}_{SM} [ \hat m^4 \ln (\hat m^2)]  + \hbox{Str}_{BSM} [ \hat m^4 \ln (\hat m^2)] \lesssim 10^{-55},
\end{equation}
while each of these terms individually is of order $\pm 30$. 
In this regard it is perhaps worth noting that numbers of the magnitude $10^{-55}$ do quite naturally show up if one considers \emph{non-perturbative} SM effects --- indeed non-perturbative effects are typically of order $e^{-1/\alpha}$ and when $\alpha$ is evaluated at the electro-weak scale $\alpha\approx 1/128$ and one has  $e^{-1/\alpha}\approx e^{-128} \approx 2.6 \times 10^{-56}$. This might only be a coincidence, but the similarity in magnitudes is certainly suggestive.

\section{Standard model before spontaneous electro-weak symmetry breaking}

Let us now consider the SM \emph{before} undergoing spontaneous electro-weak symmetry breaking. 
In the unbroken phase almost all of the SM particles are massless, except for the Higgs which has \emph{negative} (mass)$^2$. 
The $Z^0$ and $W^\pm$ merge into a SU(2) triplet $W$, and the leptons de-merge into three left-handed doublets ($g=2\times2\times 3 = 12$), plus six right-handed singlets ($g=2\times6 = 12$). 
Again I set $d=(-1)^{2S} g$. 
Note that I include 3 right-handed singlets for the tau, muon and electron, and 3 right-handed singlets for the neutrinos, this being the minimalist extension of the original standard model to include neutrino masses.
The situation is summarized in Table 2.

\begin{table}[!htp]
\begin{center}
\caption{\bf Standard model before electro-weak symmetry breaking.}
\begin{tabular}{|c|r|c|c|r|r|c|}
\hline
\hline
particle	&	$d$	&	(mass/GeV)	&$\hat m = m/m_H$	&  $d\times \hat m^2$ &	$d \times \hat m^4$	&	$d \times \hat m^4 \ln |\hat m^2|$	\\
\hline
\hline
Higgs	&	+4	&	$ (125.02)i$	&	$i$	&	$-4$	&	+4	&	0	\\
$W$	        &	+6	&	0	&	0	&	0	&	0	&	0	\\
\hline
top	        &	$-12$	&	0	&	0	&	0	&	0	&	0	\\
bottom	&	$-12$	& 0	&	0	&	0	&	0	&	0	\\
charm	&	$-12$	&0	&	0	&	0	&	0	&	0	\\
strange	&	$-12$	&0	&	0	&	0	&	0	&	0	\\
up	        &	$-12$	&0	&	0	&	0	&	0	&	0	\\
down	&	$-12$	&0	&	0	&	0	&	0	&	0	\\
\hline
gluons	&	+16	                &	0	&	0	&	0	&	0	&	0	\\
\hline
(leptons)$_L$	&	$-12$	&	0	&	0	&	0	&	0	&	0\\
(leptons)$_R$	&	$-12$       &      0	&	0	&	0	&	0	&	0	\\	
\hline
hyper-photon	&	+2	        &	0	&	0	&	0	&	0	&	0	\\
\hline
\hline
Str$_{SM}$[X]	&	$-68$ 	&	\xmark	&	\xmark	&	$-4$	&	$+4$	&	0	\\
\hline
\hline
\end{tabular}
\end{center}

\smallskip
{\bf Explanation:} Values of Str[1], Str[$\hat m^2$], Str[$\hat m^4$], and Str[$\hat m^4 \ln \hat m^2$] in the SM sector before electro-weak symmetry breaking.
\label{T:2}
\end{table}%

\vspace{-10pt}
\begin{itemize}
\item 
Note that in the standard model sector Str$_{SM}$[1] = $-68$ is unchanged, as it should be.  (Spontaneous electro-weak symmetry breaking merely moves bosonic and fermionic modes around, it does not create or destroy modes.) So Str$_{BSM}$[1] = $+68$ as previously. The BSM sector contains at least 68 bosonic degrees of freedom.
\enlargethispage{40pt}
\item
Note that Str$_{SM}$[$m^2$] = $-4 m_H^2$ and Str$_{SM}$[$m^4$] = $+4 m_H^4$ are both \emph{changed} compared to the broken phase. This is not unexpected, and actually gives us extremely useful information: Enforcing Pauli's sum rules, this implies that both  Str$_{BSM}$[$m^2$] and Str$_{BSM}$[$m^4$] must change during spontaneous electro-weak symmetry breaking, which in turn implies that at least part of the BSM spectrum must be sensitive to the onset of spontaneous electro-weak symmetry breaking, and so at least part of the BSM spectrum must couple to the Higgs. (So at least part of the BSM spectrum must be ``not entirely dark'', though it might couple only weakly to the SM sector.)
\end{itemize}
Furthermore, before spontaneous electro-weak symmetry breaking we deduce that the BSM spectrum must satisfy:
\begin{equation}
  \hbox{Str}_{BSM}[1] =  +68;   \qquad
  \hbox{Str}_{BSM}[m^2] =  4  m_H^2;  \qquad
  \hbox{Str}_{BSM}[m^4] = -4 m_H^4. 
 \label{E:bsm1-3}
\end{equation}
The BSM sector is (still) boson dominated as it should be, but now, since we know that both $\hbox{Str}_{BSM}[m^2] >0$ and $\hbox{Str}_{BSM}[m^4] <0$ in the unbroken phase,  we can deduce that there must be at least one fermion in the BSM spectrum.
Therefore there must be at least 69 bosonic degrees of freedom in the BSM sector.


\section{Discussion}

The current observational situation regarding the usual places to look for BSM physics is rather bleak.
The theoretically attractive versions of supersymmetry, technicolor, preons, large extra dimensions, strong gravity, \emph{etcetera}, are increasingly being confined to small regions of parameter space --- many might argue, to unnaturally small and undesirable regions of parameter space. 
With a lack of any direct observational evidence in favour of BSM physics, it becomes more critical to assess indirect evidence in favour of BSM physics.  

The Pauli sum rules, being based on basic symmetry principles (Lorentz invariance of the zero-point stress-energy tensor), and highly desirable phenomenology (finiteness of the zero-point stress energy tensor), are arguably a minimalist starting point.
As we have seen above, if we assume the Pauli sum rules as a minimalist starting point, then we can at least make some qualitative observations:
\begin{itemize}
\item There must be BSM physics.
\item The BSM sector must be boson dominated.
\item Parts of the BSM sector must couple to the Higgs.
\end{itemize}
While these deductions may appear weak, they are based on truly minimalist and physically well-motivated input: Lorentz invariance and/or finiteness of the zero-point stress-energy tensor. It is fascinating to see how Pauli's 67 year old observations from 1951 still resonate in the modern era.

\section*{Acknowledgments}

This research was supported by the Marsden Fund, 
through a grant administered by the Royal Society of New Zealand. 


\end{document}